\documentclass[useAMS,usenatbib]{mn2e}
\usepackage{times}
\usepackage{epsfig}
\def\beq{\begin{eqnarray}}
\def\eeq{\end{eqnarray}}

\title[Strange stars with different quark mass scalings]{Strange stars with different quark mass scalings}
\author[Ang Li, Ren-Xin Xu, and Ju-Fu Lu]{Ang Li$^{1}$\thanks{E-mail: liang@xmu.edu.cn}; Ren-Xin
Xu$^{2}$\thanks{r.x.xu@pku.edu.cn}; Ju-Fu Lu $^{1}$
\\$^{1}$Department of Physics and Institute of Theoretical Physics
and Astrophysics, Xiamen University, Xiamen 361005, China\\
$^{2}$School of Physics and State Key Laboratory of Nuclear Physics
and Technology, Peking University, Beijing 100871, China}
\begin{document}

\pagerange{\pageref{firstpage}--\pageref{lastpage}} \pubyear{2008}

\maketitle

\label{firstpage}

\begin{abstract}
We investigate the stability of strange quark matter and the
properties of the corresponding strange stars, within a wide range
of quark mass scaling. The calculation shows that the resulting
maximum mass always lies between 1.5$M_{\odot}$ and 1.8$M_{\odot}$
for all the scalings chosen here. Strange star sequences with a
linear scaling would support less gravitational mass, and a change
(increase or decrease) of the scaling around the linear scaling
would lead to a higher maximum mass. Radii invariably decrease with
the mass scaling. Then the larger the scaling, the faster the star
might spin. In addition, the variation of the scaling would cause an
order of magnitude change of the strong electric field on quark
surface, which is essential to support possible crusts of strange
stars against gravity and may then have some astrophysical
implications.
\end{abstract}


\section{Introduction}           
\label{sect:intro}

In studying the equation of state (EOS) of ordinary quark matter,
the cruial point is to treat quark confinement in a proper way.
Except the conventional bag mechanism (where quarks are
asymptotically free within a large bag), an alternative way to
obtain confinement is based on the density dependence of quark
masses, then the proper variation of quark masses with density would
mimic the strong interaction between quarks, which is the basic idea
of the quark mass-density-dependent model.

Originally, the interaction part of the quark masses was assumed to
be inversely proportional to the density~(Fowler et al. 1981;
Chakrabarty 1991; Chakrabarty 1993; Chakrabarty 1996), and this
linear scaling has been extensively applied to study the properties
of strange quark matter (SQM). However, this class of scaling is
often criticized for its absence of a convincing derivation~(Peng
2000). Then a cubic scaling was derived based on the in-medium
chiral condensates and linear confinement~(Peng 2000). and has been
widely used afterwards~(Lugones $\&$ Horvath 2003; Zheng et al.
2004; Peng et al. 2006; Wen et al. 2007; Peng et al. 2008). But this
deriving procedure is still not well justified since it took only
the first order approximation of the chiral condensates in medium.
Incorporating of higher orders of the approximation would
nontrivially complicate the quark mass formulas~(Peng 2009). In
fact, there are also other mass scalings in the literatures~(Dey et
al. 1998; Wang 2000; Zhang et al. 2001; Zhang $\&$ Su 2002; Zhang
$\&$ Su 2003).

Despite the big uncertainty of the quark mass formulas, this model,
after all, is no doubt only a crude approximation to QCD. For
example, the model may not account for quark system where realistic
quark vector interaction is non-ignorable. However, we can not get a
general idea of how the strong interaction acts from the fundamental
theory of strong interactions in hand, i.e. QCD. Until this
stimulating controversy is solved, we feel safe to take the
pragmatic point of view of using the model. This work does not claim
to answer how Nature works. However, it may shed some light on what
may happen in interesting physical situations. In this respect, the
quark mass-density-dependent model has been, and still is, an
interesting laboratory.

The aim of the present paper then, is to study in what extent this
scaling model is allowed to study the properties of SQM. To this
end, we treat the quark mass scaling as a free parameter, to
investigate the stability of SQM and the variation of the predicted
properties of the corresponding strange stars (SSs) within a wide
scaling range. Furthermore, we try to demonstrate the general
features of SSs related to astrophysics observations, whatever the
value of the free parameters.

The paper is organized as follows. In Section 2 we describe the
formalism applied in calculating the EOS of the SQM in the quark
mass-density-dependent model. In Section 3 we present the structure
of the stars made of this matter, including mass-radius relation,
spin frequency, electric properties of the quark surface. Finally in
Section 4 we address our main conclusions.

\section{{\bf The} Model}
\label{sect:Mod}

As usually done, we consider SQM as a mixture of interacting $u$,
$d$, $s$ quarks, and electrons, where the mass of the quarks $m_{q}$
($q = u, d, s$)  is parametrized with the baryon number density
$n_{\mathrm{b}}$ as follows:
\begin{equation}
m_q \equiv m_{q0}+ m_{\mathrm{I}}=m_{q0}+\frac{C}{n_{\mathrm{b}}^x},
\label{mqT0}
\end{equation}
where $C$ is a parameter to be determined by stability arguments.
The density-dependent mass $m_{q}$ includes two parts: one is the
original mass or current mass $m_{q0}$, the other is the interacting
part $m_{\mathrm{I}}$. The exponent of density $x$, i.e. the quark
mass scaling, is treated as a free parameter in this paper.

Denoting the Fermi momentum in the phase space by $\nu_i$ ($i=u, d,
s,e^-$), the particle number densities can then be expressed as
\begin{equation} \label{nimod}
n_i =g_i\int \frac{\mathrm{d}^3{\bf p}}{(2\pi\hbar)^3}
=\frac{g_i}{2\pi^2} \int_0^{\nu_i}\, p^2\,\mbox{d}p
=\frac{g_i\nu_i^3}{6\pi^2}{\bf ,}
\end{equation}
and the corresponding energy density as
\begin{equation} \label{Emod}
\varepsilon
=\sum_i\frac{g_i}{2\pi^2}\int_0^{\nu_i}\sqrt{p^2+m_i^2}\,p^2\,\mbox{d}p{\bf
.}
\end{equation}

The relevant chemical potentials $\mu_u$, $\mu_d$, $\mu_s$, and
$\mu_e$ satisfy the weak-equilibrium condition (we assume that
neutrinos leave the system freely):
\begin{eqnarray}
 \mu_u+\mu_e=\mu_d, ~~~\mu_d=\mu_s{\bf .} \label{weak}
\end{eqnarray}

For the quark flavor $i$ we have
\begin{eqnarray}
\mu_i &=& \frac{\mathrm{d} \varepsilon}{\mathrm{d} n_i}
|_{\{n_{k\neq i}\}}= \frac{\partial \varepsilon_i}{\partial \nu_i}
\frac{\mathrm{d}\nu_i}{\mathrm{d} n_i}
 +\sum_j \frac{\partial \varepsilon}{\partial m_j}\frac{\partial m_j}{\partial n_i}
\nonumber \\
&=&
  \sqrt{\nu_i^2+m_i^2}
  +\sum_j n_j\frac{\partial m_j}{\partial n_i}
   f\!\left(\frac{\nu_j}{m_j}\right),
 \label{mui}
\end{eqnarray}
where
\begin{equation}
f(a) \equiv \frac{3}{2a^3} \left[
 a\sqrt{1+a^2}-\ln\left(a+\sqrt{1+a^2}\right)
\right].
\end{equation}
We see clearly from Equ.~(\ref{mui}) that since the quark masses are
density dependent, the derivatives generate an additional term with
respect to the free Fermi gas model.

For electrons, we have
\begin{equation} \label{muevsne}
\mu_e=\sqrt{\left(3\pi^2n_e\right)^{2/3}+m_e^2}{\bf .}
\end{equation}

The pressure is then given by
\begin{eqnarray}
P&=& -\varepsilon + \sum_i \mu_i n_i
\nonumber \\
&=&
  -\Omega_0
  +\sum_{ij} n_i n_j\frac{\partial m_j}{\partial n_i}
   f\left(\frac{\nu_j}{m_j}\right)
\nonumber \\
&=& -\Omega_0
  +n_{\mathrm{b}}\frac{\mathrm{d}m_{\mathrm{I}}}{\mathrm{d}n_{\mathrm{b}}}
   \sum_{j=u,d,s} n_j\ f\!\left(\frac{\nu_j}{m_j}\right){\bf ,}
 \label{pressure}
\end{eqnarray}
with $\Omega_0$\ being the free-particle contribution:
\begin{eqnarray}
\Omega_0 &=& -\sum_i\frac{g_i}{48\pi^2} \left[
 \nu_i\sqrt{\nu_i^2+m_i^2}\left(2\nu_i^2-3m_i^2\right)
\right.
\nonumber\\
&& \phantom{-\sum_i\frac{g_i}{48\pi^2}[}
 \left.
 +3m_i^4\,\mbox{arcsinh}\left(\frac{\nu_i}{m_i}\right)
\right].
\end{eqnarray}

The baryon number density and the charge density can be given as:
\begin{equation} \label{qmeq3}
n_{\mathrm{b}}=\frac{1}{3}(n_u+n_d+n_s){\bf ,}
\end{equation}

\begin{equation} \label{qmeq4}
Q_{\mathrm{q}}=\frac{2}{3}n_u-\frac{1}{3}n_d-\frac{1}{3}n_s-n_e.
\end{equation}
The charge-neutrality condition requires $Q_{\mathrm{q}}=0$.

Solving Equs. (\ref{weak}), (\ref{qmeq3}), (\ref{qmeq4}), we can
determine $n_u$, $n_d$,  $n_s$, and $n_e$ for a given total baryon
number density $n_{\mathrm{b}}$. The other quantities are obtained
straightforwardly.

In the present model, the parameters are: the electron mass
$m_e=0.511$ MeV, the quark current masses $m_{u0}$, $m_{d0}$,
$m_{s0}$, the confinement parameter $C$ and the quark mass scaling
$x$. Although the light-quark masses are not without controversy and
remain under active investigations, they are anyway very small, and
so we simply take $m_{u0}=5$ MeV, $m_{d0}=10$ MeV. The current mass
of strange quarks is $95\pm 25$ MeV according to the latest version
of the Particle Data Group~\cite{Yao06}


\begin{figure}
\includegraphics[width=8.0cm]{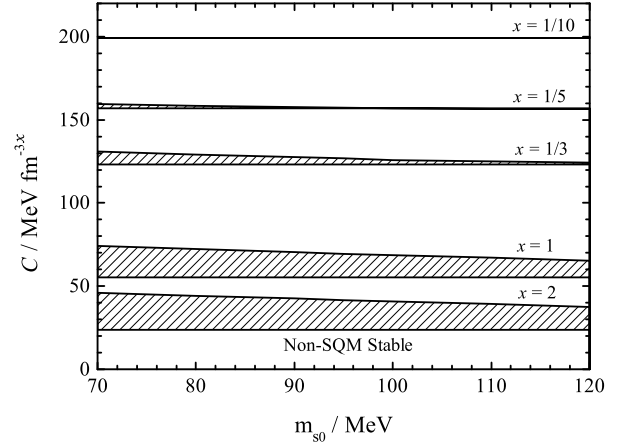}
\caption{The stability window of the SQM at zero pressure with the
quark mass scaling parameter $x = 1/10, 1/5, 1/3, 1, 2$. The
stability region (shadow region), is where the energy per particle
is lower than 930 MeV and two-flavor quark matter is unstable. }
\label{fig1}
\end{figure}

We now need to establish the conditions under which the SQM is the
true strong interaction ground state. That is, we must require, at
$P=0, E/A\leq M(^{56}{\rm Fe})c^2/56=930$ MeV for the SQM and
$E/A>930$ MeV for two-flavor quark matter (where $M(^{56}{\rm Fe})$
is the mass of $^{56}{\rm Fe}$) in order not to contradict standard
nuclear physics. The EOS will describe stable SQM only for a set of
values of ($C,m_{s0}$) satisfying these two conditions, which is
given in Fig.~1 as the ``stability window''. Only if the
$(C,m_{s0})$ pair is in the shadow region, SQM can be absolutely
stable, therefore the range of $C$ values is very narrow for a
chosen $m_{s0}$ value. As shown in Fig.~1, the allowed region
decreases for decreasing value of $x$. When $x = 1/10$ it approaches
to a very narrow area around $C$ = 199.1 MeV fm$^{{\rm -3}x}$.

We then illustrate in Fig.~2 the density dependence of
$m_{\mathrm{I}}$ with the quark mass scaling $x = 1/10, 1/3, 1, 3$.
The calculation is done with $m_{s0}=95$ MeV and $C$ values
corresponding to the upper boundaries defined in Fig. 1 (the same
hereafter), that is, the system always lies in the same binding
state (for each $x$), i.e, E/A = 930 MeV. We presented those $C$
values in the last row of the Table.~1. Clearly the quark mass
varies in a very large range from very high density region
(asymptotic freedom regime) to lower densities, where confinement
(hadrons formation) takes place. It is compared with Dey et al.'s
scaling (dash-dotted)~\cite{Dey98}.
\begin{figure}
\includegraphics[width=8.0cm]{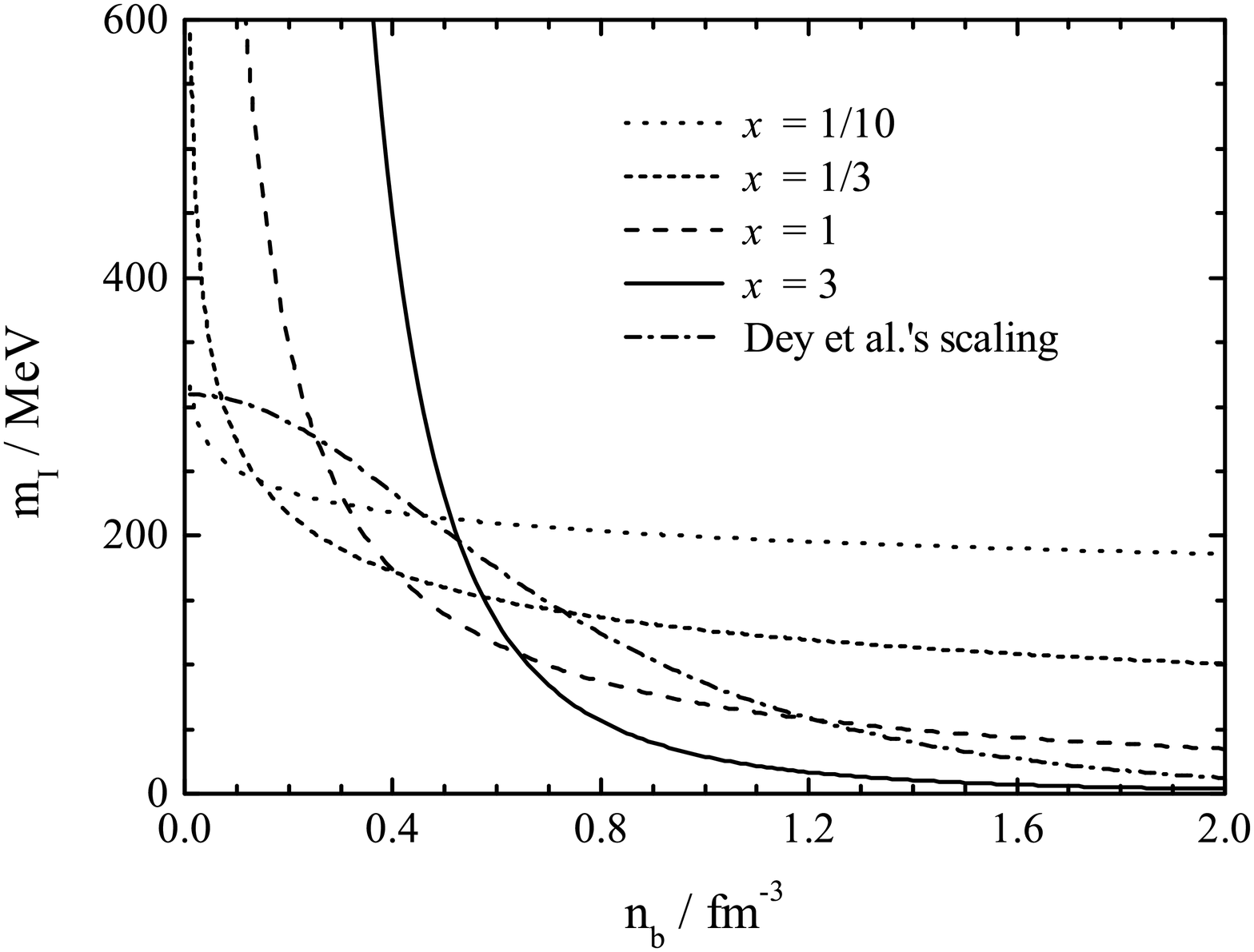}
\caption{The density dependence of $m_{\mathrm{I}}$ with the quark
mass scaling parameter $x = 1/10, 1/3, 1, 3$. The calculation is
done with $m_{s0}=95$ MeV and $C$ values presented in the last row
of the Table.~1 (see text for details). It is compared with Dey et
al.'s scaling (dash-dotted line)~(Dey et al. 1998). } \label{fig2}
\end{figure}

\section{Results and Discussion}
\label{sect:Res}

The resulting EOSs of SQM are shown in Fig.~3 for all considered
models. Because the sound velocity $v = \mid dP/d\rho\mid^{~1/2~}$
should be smaller than $c$ (velocity of light), unphysical region
excluded by this condition has been displayed with scattered dots.
For the $x$ values chosen here, they have quite different behavior
at low density, basically falling into two sequences. At small
scalings ($x = 1/10, 1/5, 1/3$) the pressure increases rather slowly
with density; while the curve turns to rapidly increase with density
at relatively large $x$ values ($x = 1, 2, 3$). They cross at
$\varepsilon \sim $ 800 MeV fm $^{-3}$, then tend to be
asymptotically linear relations at higher densities, and a larger
$x$ value leads to a stiffer EOS. Meanwhile we check also the
stability of such a quark matter, since some EOSs in Fig 3 ($x = 1,
2, 3$) are rather stiff for small pressures. We present in Fig.~4
the total pressure as a function of neutron chemical potential in
quark matter for all considered models, and comparison with that of
typical nuclear matter (obtained from the Brueckner-Hartree-Fock
approach~\cite{Li06}). We see clearly from the figure that the quark
matter tends to be more stable than nuclear matter for all
considered models.

The behavior of EOSs would be mirrored at the prediction of
mass-radius relations of the corresponding SSs, as is shown in the
Fig.~5. For the first sequence, the maximum mass occurs at a low
central density (as shown in Table.~\ref{Table1}), so a higher
maximum mass is obtained due to a stiffer EOS, and with the increase
of $x$ value, the maximum mass is reduced from 1.78$M_{\odot}$ at $x
= 1/10$ down to 1.61 $M_{\odot}$ at $x = 1/3$; While we observe a
slight increase of the maximum mass with $x$ value for the second
sequence: from 1.56$M_{\odot}$ at $x = 1$ up to 1.62 $M_{\odot}$ at
$x = 3$. Anyway the resulting maximum mass lies between
1.5$M_{\odot}$ and 1.8$M_{\odot}$ for a rather wide range of $x$
value chosen here (0.1 -- 3), which may be a pleasing feature of
this model: well-controlled. To see the region of stellar parameters
allowed by this model, we plot in Fig.~5 also the M(R) curves for
lower boundaries defined in Fig.~1 for $x = 1/5, 1/3, 1$ with grey
lines.

The radii, on the other hand, decrease invariably with $x$ value.
Employing the empirical formula connecting the maximum rotation
frequency with the maximum mass and radius of the static
configuration~\cite{Gou99}, we get also the maximum rotational
angular frequency $\Omega_{\rm max}$ as $7730{\Large \left(
\frac{M_{\odot }^{\rm stat}}{M_{\odot }} \right)^{\frac{1}{2}}\left(
\frac{R^{\rm stat}_{M_{\odot }}}{10 {\rm km}}
\right)^{-\frac{3}{2}}}$rad s$^{-1}$. As a result, a larger $x$
value results in a larger maximum spin frequency, SSs with $x = 3$
can rotate at a frequency of $f_{\rm max}$ = 2194 Hz. More detailed
results can be found in Table.~\ref{Table1}.

\begin{table}
\begin{center}
\begin{minipage}[]{85mm}
\caption{\small Calculated results for the gravitational masses,
radii, central baryon densities (normalized to the saturation
density of nuclear matter, $n_0$ = 0.17 fm$^{-3}$), and the maximum
rotational frequencies for the maximum-mass stars of each strange
star sequence. The calculation is done with $m_{s0}=95$ MeV and
 $C$ values presented in the last row of this table. } \label{Table1}
 \end{minipage}
\end{center}
\begin{center}
 \begin{tabular}{cccccccccc}
  \hline
$x$ & $1/10$  &  $1/5$ & $1/3$ & $1$ & $2$ & $3$\\
  \hline
$M/M_{\odot}$  & 1.78  & 1.66 & 1.61 & 1.56 & 1.61 & 1.62\\
$R/{\rm km}$         & 13.2  & 10.5 & 9.38 & 8.10 & 7.97 & 7.89\\
$n_c/n_0$      & 4.35  & 6.47 & 7.88 & 10.1 & 10.2 & 10.3\\
$f_{\rm max}/{\rm Hz}$    & 1066 & 1446 & 1691 & 2072 & 2159 & 2194\\
\hline
$C/{\rm MeV~fm}^{-3x}$& 199.1 & 157.2 & 126.8 & 69.5 & 41.7 & 28.8 \\
  \hline
\end{tabular}
\end{center}
\end{table}

\begin{figure}
\includegraphics[width=8.0cm]{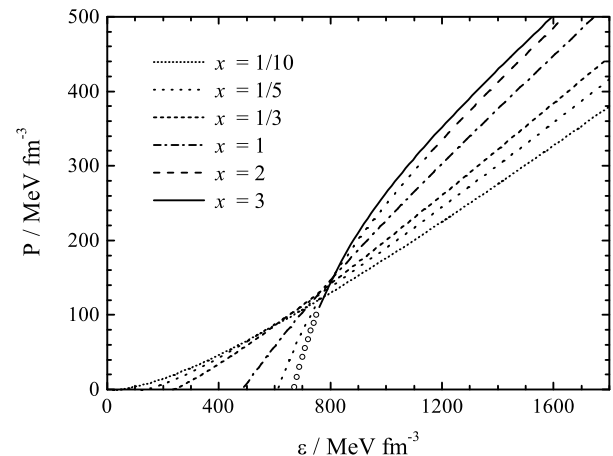}
\caption{The EOSs of SQM for all considered models. Unphysical
region excluded by \textbf{this} condition has been displayed with
scattered dots (see text for details). } \label{fig3}
\end{figure}

\begin{figure}
\includegraphics[width=8.0cm]{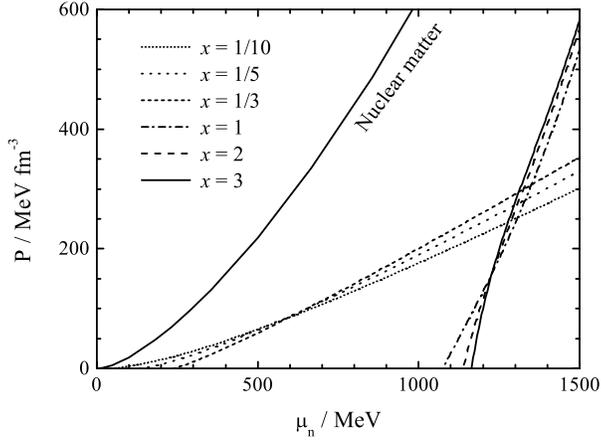}
\caption{The total pressure as a function of neutron chemical
potential in SQM for all considered models, and comparison with that
of typical nuclear matter (see text for details).} \label{fig4}
\end{figure}

\begin{figure}
\includegraphics[width=8.0cm]{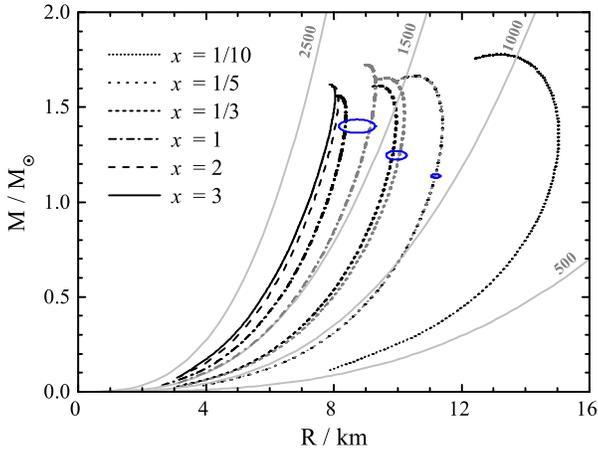}
\caption{The mass-radius relations of SSs for all considered models.
M(R) curves for lower boundaries defined in Fig.~1 with the quark
mass scaling parameter $x = 1/5, 1/3, 1$ are presented with grey
lines. Contours of the maximum rotation frequencies are given by the
light grey curves~(Gourgoulhon et al. 1999). } \label{fig5}
\end{figure}

In addition, the surface electric field could be very strong near
the bare quark surface of a strange star because of the mass
difference of the strange quark and the up (or down) quark, which
could play an important role in producing the thermal emission of
bare strange stars by the Usov mechanism~(Usov 1998; Usov 2001). The
strong electric field is also very crucial in forming a possible
crust around a strange star, which has been investigated extensively
by many authors~(for a recent development, see Zdunik et al. 2001).
Furthermore, it should be noted that this electric field may have
some important implications on pulsar radio emission mechanisms~(Xu
et al. 2001). Therefore it is very worthwhile to explore how the
mass scaling influences the surface electric field of the stars, and
possible related astronomical observations in turn may drop a hint
on what the proper mass scaling would be.

Adopting a simple Thomas-Fermi model, one gets the Poisson's
equation~(Alcock et al. 1986):
\begin{equation}
{d^2 V\over dz^2} = \left\{    \begin{array}{ll}
{4\alpha\over 3\pi}(V^3-V_{\rm q}^3) & z\leq 0,\\
{4\alpha\over 3\pi} V^3 & z > 0,
\end{array}     \right.
\end{equation}
where $z$ is the height above the quark surface, $\alpha$ is the
fine-structure constant, and $V_{\rm q}^3/(3\pi^2 \hbar^3 c^3)$ is
the quark charge density inside the quark surface.
Together with the physical boundary conditions $\{ z \rightarrow
-\infty: V \rightarrow V_{\rm q}, dV/dz \rightarrow 0;~~ z
\rightarrow +\infty: V \rightarrow 0,   dV/dz \rightarrow 0 \}$, and
the continuity of $V$ at $z=0$ requires $V(z=0) = 3V_{\rm q}/4$, the
solution for $z > 0$ finally leads to
\begin{equation}
V={3V_{\rm q}\over \sqrt{6\alpha\over\pi}V_{\rm q}z+4}~~ ({\rm
for}~z > 0).
\end{equation}
The electron charge density can be calculated as $ V^3/(3\pi^2
\hbar^3 c^3) $, therefore the number density of the electrons is
\begin{equation}
n_{\rm e}  =  {9V_{\rm q}^3\over \pi^2 (\sqrt{6\alpha\over\pi}V_{\rm
q}z+4)^3} \label{ne}
\end{equation}
and the electric field above the quark surface is finally
\begin{equation}
E = \sqrt{2\alpha\over 3\pi} \cdot {9 V_{\rm q}^2 \over
    (\sqrt{6\alpha\over \pi} V_{\rm q} \cdot z + 4)^2} \label{E}
\end{equation}
which is directed outward.

We see from Fig.~5 (take $x = 1/3$ for example) that although the
electric field near the surface is about $ 10^{18}$ V cm$^{-1}$, the
outward electric field decreases very rapidly above the quark
surface, and at $z\sim 10^{-8}$ cm, the field gets down to $\sim
5\times 10^{11}$ V cm$^{-1}$, which is of the order of the
rotation-induced electric field for a typical Goldreich-Julian
magnetosphere. To change the mass scaling mainly has two effects:
First, it affects a lot the surface electric field, and a small
scaling parameter leads to an enhanced electric field. The weakening
of electric field would be almost a order of magnitude large (from
$10^{17}$ V cm$^{-1}$ to $10^{18}$ V cm$^{-1}$), which may have some
effect on astronomical observations. Second, a larger scaling would
slow the decrease of the electric field above the quark surface.

\begin{figure}
\includegraphics[width=8.0cm]{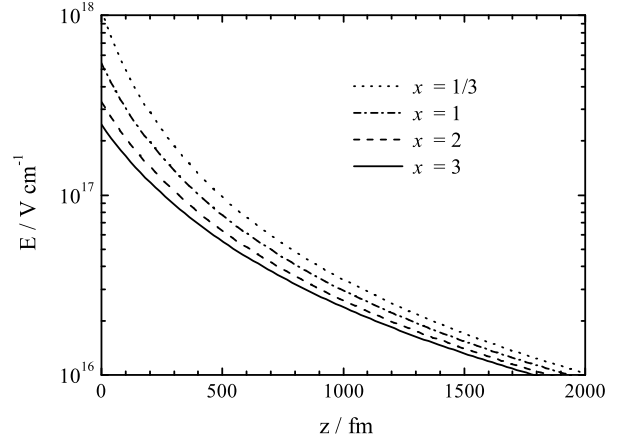}
\caption{The electric field above the quark surface with the quark
mass scaling parameter $x =  1/3, 1, 2, 3$.} \label{fig6}
\end{figure}

\section{Conclusions}
\label{sect:conclusion}

In this paper, we investigate the stability of SQM within a wide
scaling range, i.e. from 0.1 to 3. We study also the properties of
the SSs made of the matter. The calculation shows that the resulting
maximum mass always lies between 1.5$M_{\odot}$ and 1.8$M_{\odot}$
for all the mass scalings chosen here. Strange star sequences with a
linear scaling would support less gravitational mass, a change
(increase or decrease) of the scaling parameter around the linear
scaling would result in a higher maximum mass. Radii invariably
decrease with the mass scaling. Then the larger the scaling, the
faster the star rotates. In addition, the variation of the scaling
may cause an order of magnitude change of the surface electric
field, which may have some effect on astronomical observations.

\section*{Acknowledgments}

We would like to thank an anonymous referee for valuable comments
and suggestions, and acknowledge Dr. Guang-Xiong Peng for beneficial
discussions. This work was supported by the National Basic Research
Program of China under grant 2009CB824800, the National Natural
Science Foundation of China under grants 10778611, 10833002,
10973002 and the Youth Innovation Foundation of Fujian Province
under grant 2009J05013.

\end{document}